\documentclass[aps,prb,amsmath,amssymb,reprint,footinbib,superscriptaddress]{revtex4-2}

\usepackage{graphicx}
\usepackage{float}
\usepackage{bm}
\usepackage[colorlinks,allcolors=blue]{hyperref}
\usepackage{siunitx}
\usepackage{natbib}
\usepackage{bold-extra}
\usepackage[normalem]{ulem}

\begin{document}

\title{Wiedemann-Franz behavior at the Weyl points in compressively strained HgTe}

\date{\today}

\author{Abu Alex Aravindnath}
\email{abu-alex.aravindnath@physik.uni-wuerzburg.de}
\affiliation{Experimentelle Physik III, Physikalisches Institut, Universit\"{a}t W\"{u}rzburg, Am Hubland, 97074 W\"{u}rzburg, Germany}
\affiliation{Institute for Topological Insulators, Universit\"{a}t W\"{u}rzburg, Am Hubland, 97074 W\"{u}rzburg, Germany}
\affiliation{Max Planck Institute for Chemical Physics of Solids, N\"{o}thnitzer Straße 40, 01187 Dresden, Germany}
\author{Yi-Ju Ho}
\author{Fabian Schmitt}
\author{Dongyun Chen}
\author{Johannes Kleinlein}
\author{Wouter Beugeling}
\author{Hartmut Buhmann}
\author{Stanislau U. Piatrusha}
\email{stanislau.piatrusha@physik.uni-wuerzburg.de}
\affiliation{Experimentelle Physik III, Physikalisches Institut, Universit\"{a}t W\"{u}rzburg, Am Hubland, 97074 W\"{u}rzburg, Germany}
\affiliation{Institute for Topological Insulators, Universit\"{a}t W\"{u}rzburg, Am Hubland, 97074 W\"{u}rzburg, Germany}
\author{Laurens W. Molenkamp}
\email{laurens.molenkamp@physik.uni-wuerzburg.de}
\affiliation{Experimentelle Physik III, Physikalisches Institut, Universit\"{a}t W\"{u}rzburg, Am Hubland, 97074 W\"{u}rzburg, Germany}
\affiliation{Institute for Topological Insulators, Universit\"{a}t W\"{u}rzburg, Am Hubland, 97074 W\"{u}rzburg, Germany}
\affiliation{Max Planck Institute for Chemical Physics of Solids, N\"{o}thnitzer Straße 40, 01187 Dresden, Germany}

\begin{abstract}
Weyl semimetals, with their unique electronic band structure, have drawn significant interest for their potential to explore quantum anomalies in condensed matter systems. In this study, we investigate the large positive magneto-thermal conductance associated with the gravitational anomaly – one of the predicted anomalies – for a Weyl semimetal based on a compressively strained HgTe layer. We clearly identify the Weyl regime in our device and accurately extract the thermal conductance by performing thermometry measurements at liquid helium temperatures using fully electronic methods. We observe the anticipated increase in thermal conductance, and it perfectly matches the electrical conductance according to the Wiedemann-Franz law. This finding indicates that, despite the unique electronic spectrum of Weyl semimetals, the mechanism governing heat transport in this system is the same as that for electrical transport, with no additional violations of conservation laws.
\end{abstract}

\maketitle


The ability of electrons to carry heat provides a unique approach for exploring mesoscopic systems. Hence, significant interest has been recently attracted to the electron thermal transport in Weyl semimetals, characterized by having their electronic excitations governed by the Weyl equations~\cite{RevModPhys.90.015001}. These equations feature a linear band dispersion with relativistic electrodynamic properties, reminiscent of the ones encountered in particle physics~\cite{NIELSEN1983389}. This has led to Weyl semimetals being proposed as a condensed matter platform for studying phenomena analogous to those predicted in high-energy physics, such as quantum anomalies. In particular, for Weyl semimetals an increase of electrical conductance with magnetic field has been connected to the chiral anomaly~\cite{PhysRevB.88.104412,PhysRevB.91.245157}, while a positive magneto-thermal conductance has been predicted~\cite{PhysRevB.85.184503,doi:10.1073/pnas.1608881113} and attributed to the gravitational anomaly leading to the violation of separate conservation laws for energy-momentum tensors within chiral Weyl cones.

However, several challenges complicate the observation of these anomalies. One significant obstacle is that the Fermi level of available materials often lies far from the Weyl point, making it difficult to attribute positive magneto-conductance exclusively to the chiral anomaly, as it can also arise from various other mechanisms~\cite{PhysRevB.92.075205,PhysRevLett.120.026601}. Observing the gravitational anomaly is even more challenging because heat, unlike charge, is not conserved and can easily dissipate from the electron subsystem into the crystal lattice~\cite{RevModPhys.78.217}. Earlier experiments~\cite{Gooth2017,Vu2021,PhysRevB.108.L161106} have been carried out at relatively high lattice temperatures, where the electron transport is significantly influenced by interaction with phonons, leading to phonon drag and rapid electron-phonon relaxation. Furthermore, exploring thermoelectric phenomena requires using the correct set of constitutive equations~\cite{BENENTI20171,HvanHouten_1992} that reflect the experimental observables (voltage and heat flow). In contrast, the sets employed in~\cite{Gooth2017} and~\cite{Vu2021} do not meet this criterion. The issue is particularly evident in~\cite{Vu2021}, where the equations chosen lead to an incorrect sign of thermopower correction to the heat conductance. The constitutive equations for thermoelectric coefficients are often derived without directly referencing experimentally observable quantities, notably as in~\cite{ashcroft1976solid}, which is widely regarded as a primary reference in this field. However, it is essential to formulate these coefficients using directly measurable quantities to accurately capture their physical meaning. An experiment cannot measure short-circuited thermal conductance, and in an open circuit, there is always a mixture of the thermopower to the thermal conductance~\cite{HvanHouten_1992}.

\begin{figure*}
	\includegraphics[scale=0.95]{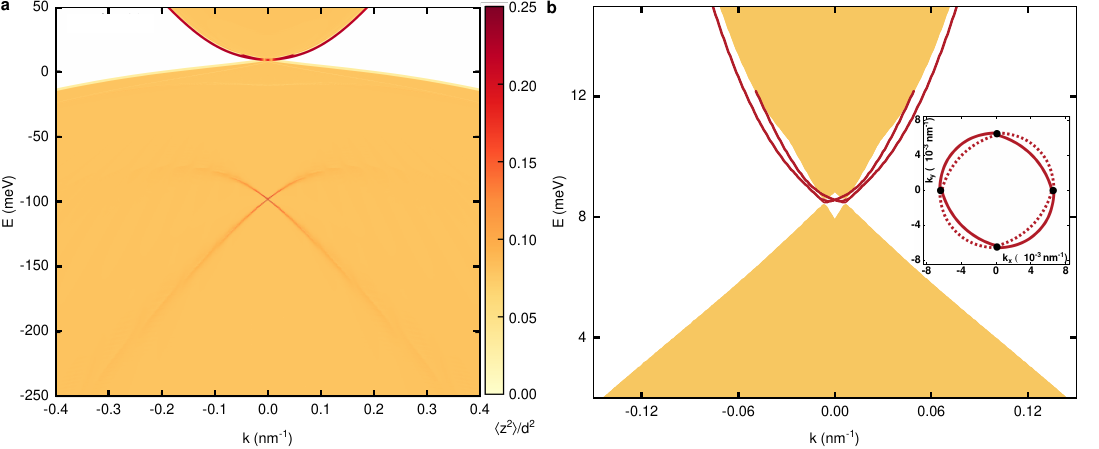}
	\caption{
		\textbf{Band structure of compressively strained HgTe}. (a) Energy momentum dispersion of HgTe under a compressive strain of 0.28\%, calculated using $\boldsymbol{k \cdot p}$ theory, with momentum $k$ along the (100) direction. In order to analyze the surface states, we use a finite-slab geometry, with surfaces perpendicular to the growth direction (001). Bulk bands are shown in orange while surface states are depicted in maroon. The massless $n$-type surface states are located near the Weyl points, with the surface Dirac cone situated deep within the valence band. (b) Magnification of (a) near the Fermi level, showing the two Weyl points on the $k_x$ axis with equal chirality and the four $n$-type surface states (maroon). The surface states are two-fold degenerate on the $k_x$ axis, making up for a total of four surface states, i.e., two at each surface. The inner pair merges into the conduction band for $k\approx 0.05\, \mathrm{nm}^{-1}$. The inset depicts the Fermi surface at the Fermi level for the top
		(solid) and bottom (dashed) surface states. These surface states at the
		Fermi level are known as ``Fermi arcs". The black dots mark the positions
		of the Weyl points. }
	\label{fig:band structure}
\end{figure*}

We previously have studied the chiral anomaly, a negative magnetoresistance at the Weyl points in a compressively strained HgTe layer~\cite{PhysRevX.9.031034}. These layers have a low intrinsic carrier density and the Fermi level can be tuned (by means of a gate voltage) from the valence band into the conduction band, enabling direct access to the Weyl points, massive and massless (i.e. topological) Volkov-Pankratov surface states, and bulk bands within the same device. In the present paper, we use this ability to unequivocally identify the Weyl regime and address the gravitational anomaly via thermal conductance measurements. These transport phenomena are fundamentally due to the linear dispersion near the Weyl points, as illustrated by the band structure in Fig.~\ref{fig:band structure} for compressively strained HgTe (calculated using $\boldsymbol{k \cdot p}$ theory \cite{10.21468/SciPostPhysCodeb.47}). The presence of topological surface states emanating from the Weyl nodes is a common property of many Weyl semimetals. The topological surface states are present because of the band crossings needed to create the Weyl points, and absent only when crystalline symmetry avoids their occurrence. 

To detect the electronic thermal conductance exclusively and avoid a strong contribution of a phonon drag component, we aim for thermal separation between the electron and phonon subsystems and cool the device to a lattice temperature of $1.33\,\unit{\kelvin}$. At this temperature, electrons primarily transfer heat to the lattice by emitting acoustic phonons, leading to a much lower electron-phonon relaxation compared to higher temperatures. Additionally, we utilize all-electronic methods of temperature measurement (via electronic noise)~\cite{KOBAYASHI2016PJA9207B-02} and heating (using a heater with hot electron diffusion)~\cite{PhysRevLett.65.1052,PhysRevLett.68.3765}.

As expected for a Weyl semimetal, we observe that the thermal conductance increases with applied in-plane magnetic field. When we compare the thermal and electrical conductance, we find them matching according to the fundamental Wiedemann-Franz law for all relevant Fermi level positions. This indicates that no anomalous physics is associated with the heat transport in this system. We discuss the implications of our result for the observation of quantum anomalies in Weyl semimetals.

\section*{Results}

\begin{figure*}
	\includegraphics[scale=0.95]{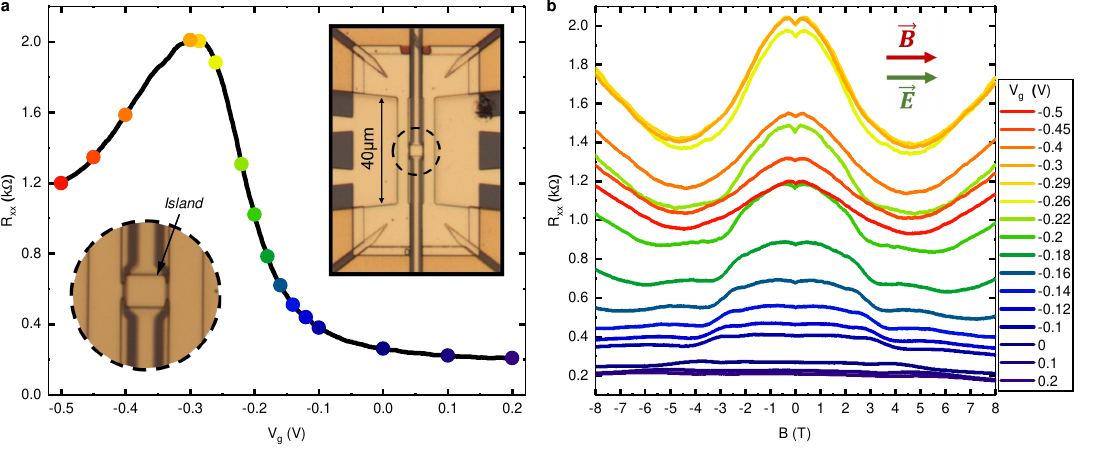}
	\caption{
		\textbf{In-plane magnetoresistance of the island}. (a) $R_\mathrm{xx}$ vs $V_\mathrm{g,i}$ at the island at $B=0$. The insets show the optical image of the device and an enlarged view of the island. (b) In-plane magnetoresistance for different positions of the island Fermi level. 
	}
	\label{fig:resistance}
\end{figure*}

For our experiment, we utilize a device shaped into a symmetric H-bar structure~\cite{PhysRevB.70.241301} (see inset of Fig.~\ref{fig:resistance}a). The two $40\,\unit{\micro\meter}$ long and $4\,\unit{\micro\meter}$ wide channels are used as heater and detector. In the centre, the channels are connected by a $4.5\,\unit{\micro\meter}$ wide and $6\,\unit{\micro\meter}$ long \textit{island} of which the transport properties are probed. Both channels and the island are equipped with separate electrostatic gates, allowing for independent control of the carrier density. Unless specified otherwise, we gate the heater and the detector channels into the conduction band, while the gate voltage at the island $V_\mathrm{g,i}$ is varied to study transport at different positions of the Fermi level.

Fig.~\ref{fig:resistance}a shows the zero-field longitudinal resistance, $R_\mathrm{xx}$(B=0T) of the island at lattice temperature $T_0\approx1.33\,\unit{\kelvin}$, as the Fermi level is tuned from the conduction band to the valence band. Weyl points are located near the resistance maximum of the gate sweep, corresponding to the charge neutrality point in the band structure. We characterize the island resistance $R_\mathrm{xx}(B)$ by measuring it as a function of in-plane magnetic field $B$ (applied parallel to the direction of heat transport in the island) at different island gate voltages $V_\mathrm{g,i}$ using lock-in measurement techniques (Methods) in Fig.~\ref{fig:resistance}b. As shown in Fig.~\ref{fig:resistance}b, we observe a negative magnetoresistance below $5\,\unit{\tesla}$~\cite{PhysRevX.9.031034} for a wide range of island gate voltages $V_\mathrm{g,i}$. The maximum effect is observed around $V_\mathrm{g,i}= -0.29\,\unit{\volt}$, where $R_\mathrm{xx}(0)\approx2\,\unit{\kilo\ohm}$ also peaks as a function of $V_\mathrm{g,i}$ [see Fig.~\ref{fig:resistance}a for gate voltage dependence of $R_\mathrm{xx}(0)$]. As discussed extensively in Ref.~\cite{PhysRevX.9.031034}, the observed negative magnetoresistance is attributed to the linear band crossings around the Weyl points (i.e., the chiral anomaly), while the observed small resistance increase up to $0.5\,\unit{\tesla}$ is related to weak antilocalization.

To study the thermal transport in our Weyl material, we use current heating, running a DC current $I_\mathrm{heat}$ through the heater channel, which increases the electron temperature at its centre to $T_\mathrm{hot}$, above the lattice temperature $T_0$. This temperature increase results in a heat flow, with rate $Q$, through the island into the detector channel. This heat flow subsequently raises the electron temperature at the detector side, $T_\mathrm{cold}$.

\begin{figure*}
	\centering
	\includegraphics[scale=0.8]{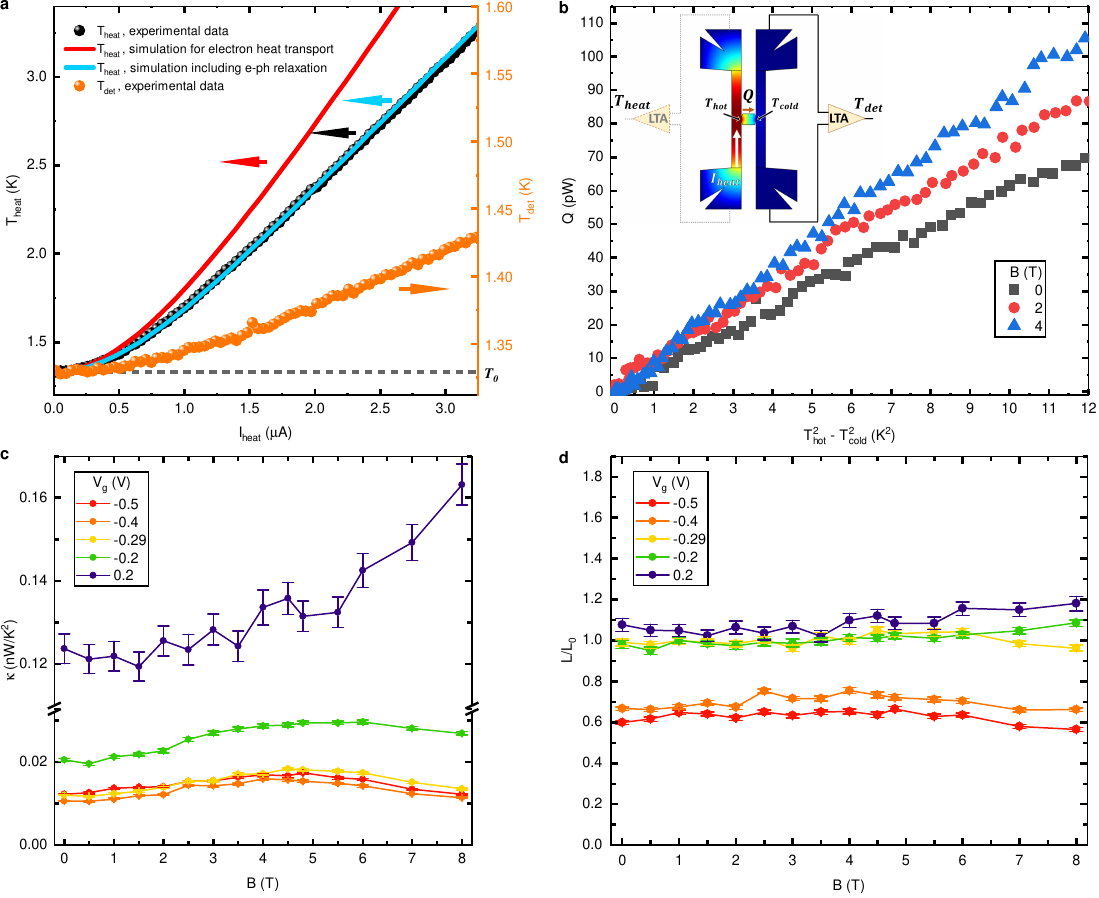}
	\caption{
		a) The average heater channel ($T_\mathrm{heat}$, black dots, left axis) and detector channel ($T_\mathrm{det}$, orange dots, right axis) temperatures, measured via noise thermometry, as a function of DC heating current, $I_\mathrm{heat}$ for in-plane magnetic field $B=0$. The Fermi level of the island is tuned close to the Weyl point at gate voltage $V_\mathrm{g,i}=-0.29\,\unit{\volt}$. The solid lines are simulations of the heater performance for pure hot electron diffusion (red) and including electron-acoustic phonon relaxation ($q_\mathrm{ph}=\Sigma_\mathrm{ph}(T^3 - T_0^3)$ with coefficient $\Sigma_\mathrm{ph}\approx0.19\,\unit{\watt\meter^{-2}\kelvin^{-3}}$) (light blue).
		b) The heat flow rate $Q$ through the island as a function of $T_\mathrm{hot}^2 - T_\mathrm{cold}^2$ for different in-plane magnetic fields: $B=0$, $2\,\unit{\tesla}$ and $4\,\unit{\tesla}$. The inset shows a schematic of the thermal transport measurement configuration with the temperature profiles in the channels. Owing to the symmetry of the device, $T_\mathrm{heat}$ is measured via running the current through the detector channel, without a second amplifier.
		c) Thermal conductance coefficient $\kappa$ of the island as a function of in-plane magnetic field for different positions of Fermi level, ranging from $n$-type regime, through the Weyl point at gate voltage $V_\mathrm{g,i}\approx-0.29\,\unit{\volt}$, and all the way to the $p$-region.
		d) Data from (c) and Fig.\ref{fig:resistance}b, combined yield the Lorenz ratio $L=\kappa / G$ ($G$ is the electrical conductance) in units of $L_0\approx2.44\times10^{-8}\,\unit{\watt\ohm\kelvin^{-2}}$. The error bars are derived from the statistical analysis of noise measurements, representing a 95\% confidence level.
	}
	\label{fig:kappa}
\end{figure*}

To determine $T_\mathrm{cold}$ (and $T_\mathrm{hot}$) we measure the power spectral density of the voltage fluctuations $S_\mathrm{V}$ of the detector channel using a conventional technique with resonant coupling to a cryogenic amplifier~\cite{KOBAYASHI2016PJA9207B-02} (Methods). By carefully designing the electrical circuit we ensure that no other parts of the sample contribute to $S_\mathrm{V}$, also excluding the potential influence of non-equilibrium noise due to the temperature difference~\cite{Tikhonov2016,Lumbroso2018}. This allows the extraction of an average electronic temperature of the detector $T_\mathrm{det}$ via the Johnson-Nyquist noise expression $S_\mathrm{V}=4k_\mathrm{B}T_\mathrm{det}R_\mathrm{ch}$~\cite{Kogan_1996}, where $R_\mathrm{ch}$ is the resistance of the channel (Fig.~\ref{fig:kappa}a). Using the symmetry of the H-bar, we can obtain the average heater temperature $T_\mathrm{heat}$ without needing to install and calibrate a second cryogenic amplifier by changing the configuration and passing the current through the detector channel instead (Fig.~\ref{fig:kappa}a), while ensuring that $R_\mathrm{ch}$ is the same for heater and detector channels.

Assuming linear response, the electron thermal conductance $K$ increases linearly with temperature as $K={\kappa}T$, where $\kappa$ is the temperature-independent thermal conductance coefficient. The heat transport equation through the island can then be written as (this relation is valid for both coherent and incoherent conductors):
\begin{equation}
	Q=\kappa/2(T_\mathrm{hot}^2-T_\mathrm{cold}^2),
	\label{eq:heat_transport}
\end{equation}
where $T_\mathrm{hot}$ and $T_\mathrm{cold}$ are the electron temperatures at the entrance and exit of the island respectively.
Since the dominant mechanism of heat loss in the channels is simply through electron diffusion to the leads, the temperatures inside the channels are not uniform but rather develop profiles towards the leads. $T_\mathrm{hot}$ and $T_\mathrm{cold}$ are thus not identical to directly measured $T_\mathrm{heat}$ and $T_\mathrm{det}$ respectively, but can still be directly determined from them, as follows.

An analytical form for the relations between centre and average temperatures ($T_\mathrm{hot}/T_\mathrm{heat}$ and $T_\mathrm{cold}/T_\mathrm{det}$) exists only for hot electron diffusion in a uniform strip. To improve the measurement precision, we go one step further, numerically model the heater temperature profile in our exact device geometry, and compare it with the measured signal (Methods). From analyzing the heater performance for larger $T_\mathrm{heat}$ range, we find that even at the low lattice temperature we use, electron-phonon relaxation influences the heat transport in the channels and identify the rate of heat loss to the phonons per unit area to be of the form $q_\mathrm{ph}=\Sigma_\mathrm{ph}(T^3 - T_0^3)$~\cite{Ridley_1991,PhysRevB.96.245417} with coefficient $\Sigma_\mathrm{ph}\approx0.19\,\unit{\watt\meter^{-2}\kelvin^{-3}}$ (Fig.~\ref{fig:kappa}a).

As a last ingredient, we find the heat flow $Q$ taking advantage of the symmetric layout and similar heat relaxation mechanism in the heater and detector. When applying the current $I_\mathrm{heat}$ to the heater, most of the Joule power $P_\mathrm{J}=I_\mathrm{heat}^2R_\mathrm{ch}$ is dissipated directly on the heater side, with only a small fraction transported to the detector as $Q$. This is obvious from Fig.~\ref{fig:kappa}a, where we observe that the temperature increase $T_\mathrm{heat}-T_0$ is more than ten times larger than $T_\mathrm{det}-T_0$. With the heat relaxation rate $P_\mathrm{J}(T_\mathrm{heat})$ in the heater channel, we can then determine $Q$ at given $T_\mathrm{det}$ as $Q\approx(2/3)P_\mathrm{J}(T_\mathrm{heat}=T_\mathrm{det})$. The coefficient $2/3$ relates to the difference between the centered and uniform heating of a diffusive system~\cite{PhysRevB.104.L161403}. For our analysis, we again account for the actual device geometry and extract the proportionality coefficient from a numerical model (Methods). Using Eq.(\ref{eq:heat_transport}) and the reasoning described above, we plot the measured values of $Q$ vs $T_\mathrm{hot}^2-T_\mathrm{cold}^2$ (Fig.~\ref{fig:kappa}b). We indeed observe that this dependence is linear, proving that the thermal conductance increases linearly with temperature. For further analysis, we focus on the thermal conductance coefficient $\kappa=K/T$.

\begin{figure*}
	\includegraphics[scale=0.95]{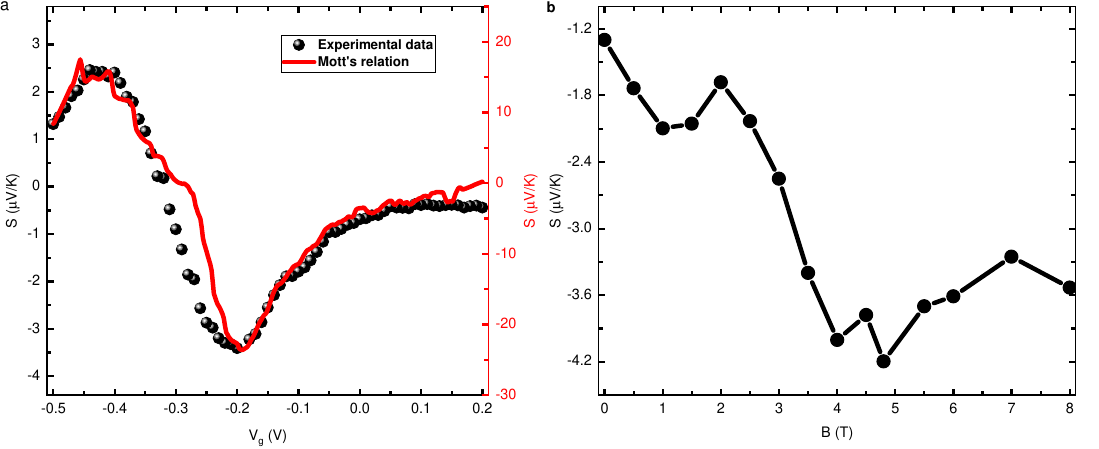}
	\caption{
		a) The thermopower of the island is measured as a function of the island gate voltage. The Fermi level within the island is adjusted from the $n$-type region ($0.2\,\unit{\volt}$) to the $p$-type region ($-0.5\,\unit{\volt}$), passing through the Weyl point at $\approx-0.29\,\unit{\volt}$. b) The thermopower of the island is investigated as a function of the in-plane magnetic field when the island is near the Weyl point.
	}
	\label{fig:thermopower}
\end{figure*}

In Fig.~\ref{fig:kappa}c the extracted $\kappa$ is shown for different positions of the island Fermi level from $n$-type to $p$-type through the Weyl point. We observe the largest $\kappa$ in the $n$-region (for gate voltage $V_\mathrm{g,i}=0.2\,\unit{\volt}$), where it exhibits a slight monotonous increase with magnetic field. Closer to the Weyl point ($V_\mathrm{g,i}=-0.2\,\unit{\volt}$ and $V_\mathrm{g,i}=-0.29\,\unit{\volt}$) $\kappa$ is more than 6 times smaller and non-monotonous, following $1/R_\mathrm{xx}$ in its behavior. Perhaps surprisingly, when moving into the $p$-region ($V_\mathrm{g,i}=-0.4\,\unit{\volt}$ and $V_\mathrm{g,i}=-0.5\,\unit{\volt}$) the thermal conductance looks nearly the same as at the Weyl point. We will argue below, that this is simply due to the enhanced hole-phonon scattering due to the large effective mass of the holes in the valence band of HgTe.

In a non-interacting fermion system, charge and heat transport are interconnected, as described by the fundamental Wiedemann-Franz law, ${\kappa}T=GL_0T$~\cite{BENENTI20171}. This law relates the electrical and thermal conductance through the Lorenz number $L_0$, which for electrons with charge $e$ is given by $L_0=(\pi^2/3)(k_\mathrm{B}/e)^2\approx2.44\times10^{-8}\,\unit{\,\watt\,\ohm\,\kelvin^{-2}}$. To check if the Wiedemann-Franz law is obeyed for our Weyl semimetal, in Fig.~\ref{fig:kappa}d we show the Lorenz ratio $L=\kappa / G$ with $G=1/R_\mathrm{xx}$ taken from the corresponding traces in Fig.~\ref{fig:resistance}b. For any $V_\mathrm{g,i}$ we observe that the variations in magneto-thermal conductance match the variations in magneto-electrical conductance, leading to an almost constant $L$ across the entire studied field range. Importantly, $L\approx{L_0}$ both at the Weyl point and in the $n$-conducting region. The slight increase of $L$ above $L_0$ at $V_\mathrm{g,i}=0.2\,\unit{\volt}$ is well understood since at this gate voltage the measured island resistance is overestimated due to the comparable sheet resistivities of island and channels.

At the same time, we observe $L$ to be significantly lower than $L_0$ in the $p$-regime at $V_\mathrm{g,i}=-0.4\,\unit{\volt}$ and $V_\mathrm{g,i}=-0.5\,\unit{\volt}$. We show in section S6 of the supplementary material, that in the $p$-regime the heat loss to the phonons in the island is almost two times larger than both for $n$-regime and close to the Weyl point. This accounts for extra heat loss and reduces the measured $L$. A strong hole-phonon relaxation is expected for HgTe, given the high effective mass of holes leading to much lower mobility and consequently increased scattering.

According to Onsager’s generalized transport equations, the presence of thermopower $S$ modifies the thermal conductance in a circuit with no net electric current so that~\cite{BENENTI20171, HvanHouten_1992, PhysRevB.95.125206}
\begin{equation}
	\kappa=\kappa_0 - S^2G.
	\label{eq:thermopower}
\end{equation}
Here $\kappa_0$ is the intrinsic thermal conductance associated with short-circuited configuration. We note that the effect described by Eq.(\ref{eq:thermopower}), does not alter the mechanism of heat transport by individual particles but rather describes the macroscopic effect of the presence of a thermoelectric voltage on heat transport, so that the Wiedemann-Franz law should hold for $\kappa_0$. For a proper analysis of $\kappa$, the contribution $S^2G$ has to be accounted for.

To address this, we measure the development of thermopower across the island with gate voltage using a second harmonic technique (see Methods). The sign-changing behavior in Fig.~\ref{fig:thermopower}a is in good agreement with the Mott relation for low temperatures $S\approx-L_0eT\frac{1}{G}\frac{dG}{d\varepsilon}\vert_{\varepsilon=\varepsilon_\mathrm{F}}$, where the derivative of the experimentally measured conductance with energy $\varepsilon$ is taken at the Fermi level $\varepsilon_\mathrm{F}$. The observed deviation in magnitude of the experimental value for $S$ is not surprising, as the exact expression for $S$ requires separately accounting for the contributions from all bands~\cite{PhysRevB.95.125206}.

Fig.~\ref{fig:thermopower}b depicts the variation of thermopower with the in-plane magnetic field $B$ at $V_\mathrm{g,i}=-0.29\,\unit{\volt}$. Despite a slight increase with $B$, the contribution $S^2G$ remains at least three orders of magnitude smaller than the measured thermal conductance coefficient $\kappa$, which thus does not require extra correction.

\section*{Discussion}

Weyl semimetals are of significant interest due to their potential for studying quantum anomalies, which play crucial role in understanding exotic transport phenomena. Although negative magnetoresistance is frequently observed in these systems, it can result from any linear crossings in the band structure close to the Fermi energy and does not necessarily serve as definitive evidence of chiral anomaly~\cite{PhysRevB.92.075205}. In contrast, thermal conductance measurements may provide a more robust indication of the non-trivial transport regime associated with quantum anomalies. For gravitational anomalies, a violation of the Wiedemann-Franz law is expected, which requires more complex assumptions about the system's transport properties, such as those described by the hydrodynamic theory of Lucas et al.~\cite{doi:10.1073/pnas.1608881113} Although the hydrodynamic model predicts Wiedemann-Franz behavior ($L=L_0$) for weak intervalley scattering and at low temperatures, it also predicts a different sign for the Seebeck coefficient $S$, thereby violating the Mott relation and not aligning with our experimental observations.

Andreev and Spivak~\cite{PhysRevLett.120.026601} derived a full set of thermoelectric coefficients of both topological and conventional conductors based on the relevant relaxation times. It predicts both Wiedemann-Franz law and Mott relation obeyed for Weyl semimetal at low temperatures, as all transport phenomena in this limit are related to the chiral anomaly. This agrees with our observations, though it does not suggest any distinct physics, as such behavior could simply stem from band crossing effects. However, in the hydrodynamic regime the authors again rather expect both relations to be not valid. Given that the specific transport regimes, particularly the hydrodynamic regime described in~\cite{doi:10.1073/pnas.1608881113} and~\cite{PhysRevLett.120.026601}, do not appear to be realized in our system, it remains open to question whether the gravitational anomaly is applicable to the experiments discussed here.

A number of experimental studies have been conducted previously to investigate quantum anomalies in Weyl semimetals. A study featuring the Weyl semimetal NbP~\cite{Gooth2017} investigates a concept called ``thermoelectric conductance", which is not directly experimentally accessible and can be rather expressed as a product of the Seebeck coefficient and the electrical conductance, $S{\cdot}G$. Since $S$ is associated with a different coefficient in the constitutive equations than thermal transport, a separate verification of the Mott relation is necessary to identify a non-trivial thermal transport regime. Another study~\cite{Vu2021} considers the Weyl phase of Bi$_{1-x}$Sb$_x$, where adherence to the Wiedemann-Franz law has been reported, similar to our results. We note that both experiments~\cite{Gooth2017,Vu2021} have been performed without experimental control of the Fermi level position and in a regime with strongly coupled electron and phonon subsystems using lattice heating at high temperatures. The validity of the Wiedemann-Franz law in this limit for the electronic component of thermal conductance is trivial and does not indicate any protected thermal transport, unlike proposed in~\cite{Vu2021}, since strongly coupled electron and phonon subsystems will reach the same local temperature and the heat flows between them are balanced~\cite{RevModPhys.93.041001}.

The Wiedemann-Franz law is generally robust as it is not based on any major approximations other than the description of transport in terms of fermionic quasiparticles. For this law to be violated (other than by the effect of Eq.(\ref{eq:thermopower})), a strong electron-electron interaction is required. While increasing the temperature might seem like a viable option to intensify electron-electron scattering, this approach is very limited. As temperature rises, electron-phonon relaxation also becomes more prominent, narrowing the viable temperature range for thermal transport experiments.
	
In graphene, the electron-phonon relaxation is weakened due to the high optical phonon energy and the small interaction with acoustic phonons~\cite{PhysRevB.93.075410}, allowing for observable deviations from the Wiedemann-Franz law at zero field due to hydrodynamic carrier flow~\cite{Waissman2022}. However, this does not apply to Weyl materials, making the feasibility of observing any transport anomalies at high temperatures questionable. Our experiment, on the other hand, demonstrates that at low temperatures, any possible special mechanisms of heat transport are not active.

We conclude that in a Weyl semimetal based on a compressively strained HgTe layer we observe an increase of electrical and thermal conductances with in-plane magnetic field matching the direction of electric field or temperature difference respectively. However, these conductances are still perfectly matched by the Wiedemann-Franz law, and the Seebeck coefficient remains in accordance with the Mott relation. These observations are in stark contrast with theories predicting a gravitational anomaly, and rather point towards a weak interaction between electrons at different Weyl nodes. The fact that our experiment is performed at low temperature, where little phonon scattering occurs, ensures that the intrinsic carrier behavior is not influenced by phonon drag, which is otherwise a very common phenomenon in semiconductor materials. At the same time, we gathered a complete set of transport coefficients for thermoelectric transport in a Weyl semimetal, which can serve as a basis for development of a transport theory in such materials.

\bibliography{manuscript.bib}

\clearpage
\newpage
\section*{Methods}

\subsection*{Material and fabrication}

The device is fabricated on a 72 nm thick HgTe layer sandwiched between two $15.2\,\unit{\nano\meter}$ top and bottom (Hg, Cd)Te layers on a GaAs substrate, grown by molecular beam epitaxy (MBE). A CdTe/ZnTe virtual substrate applies a $0.28\,\%$ compressive strain, lifting the $\Gamma_8$ band degeneracy at the $\Gamma$ point of the Brillouin zone to form Weyl points~\cite{PhysRevX.9.031034}.

The H-bar device consists of two channels connected by a central island. The channels are $40\,\unit{\micro\meter}$ long and $4\,\unit{\micro\meter}$ wide, while the island measures $6\,\unit{\micro\meter}$ in length and $4.5\,\unit{\micro\meter}$ in width. The device is patterned using electron beam and optical lithography in conjunction with wet etching. Ohmic contacts to the device are fabricated by depositing AuGe and Au layers, each with a thickness of 80nm. The top gate electrodes are composed of 5 nm Ti and 100 nm Au, with a {$15\,\unit{\nano\meter}$} HfO\textsubscript{2} gate insulator, grown via atomic layer deposition (ALD), separating them from the mesa.

\subsection*{Transport measurements}

The device is cooled in exchange gas thermally connected to a pumped $^{4}\mathrm{He}$ cryostat, resulting in base temperature $T_0\approx1.33\,\unit{\kelvin}$. The copper chip carrier is attached to a copper sample receptacle, resulting in a large effective cooling area and stable phonon temperature. In close proximity of the sample, a calibrated resistance thermometer is attached, allowing to measure the lattice temperature.

The device resistance is measured using a standard lock-in technique, wherein an alternating current with a frequency of $\approx3\,\unit{\hertz}$ and an amplitude of less than $60\,\unit{\micro\volt}$ is applied across the device and a $100\,\unit{\kilo\ohm}$ series resistor, across which the current is measured. The resistance of the island is measured using a four-terminal configuration, with the legs of the H-bar serving as the current and voltage probes. Magnetic field dependent properties are measured by applying a magnetic field at a sweeping rate of $0.1\,\unit{\tesla/\minute}$, aligned with the direction of heat and charge flow in the island and within the device plane.

\subsection*{Noise thermometry}

We measure the power spectral density $S_\mathrm{V}$ from the voltage fluctuations picked up from the detector channel using a home-built low-temperature amplifier with a voltage gain $A_\mathrm{LT}\approx5$ (the gain is determined by calibration of the measurement circuit using its thermal noise, the long-term stability is ${\delta}A_\mathrm{LT}<1\%$), connected to one side of the detector channel through a coaxial cable. The amplifier is located outside of the sample chamber, submerged in a pumped liquid helium bath, away from the magnetic field. An air-core superconducting inductor is connected in parallel to the amplifier input, resulting in combination with the coaxial cable capacitance in resonance impedance matching at $f_\mathrm{res}\approx5.7\,\unit{\mega\hertz}$. The other side of the detector channel is effectively grounded at $f_\mathrm{res}$ using a $10\,\unit{\nano\farad}$ capacitor. The remaining lines are choked for high frequency signals using $50\,\unit{\kilo\ohm}$ resistors, still allowing to apply and measure DC signals (see the full measurement schematic in Fig. S1 of the Supplemental Materials for details). This circuit results in a minimal pickup of the electronic noise from the heater side and the island itself, effectively measuring the noise temperature of the detector only.

The amplified voltage noise is guided outside of the low-temperature compartment to a room temperature amplifier with a voltage gain $A_\mathrm{RT}\approx200$ and finally into a spectrum analyzer. The spectrum analyzer measures the average square voltage $V_{SA} = \sqrt{\langle V^2\rangle}$ around $f_{res}$ in a 30\,kHz bandwidth ($\Delta f$) defined with a Gaussian filter. We convert $V_{SA}$ to spectral power density of the voltage fluctuations $S_V$ using $S_V = V_{SA}^2 /( \Delta fA_{LT}A_{RT})$, where $A_{RT}$ represents the voltage gain of room temperature amplifier.

\subsection*{Calibration of the heater and detector}

We analyze the heater performance as a function of $I_\mathrm{heat}$ up to a $T_\mathrm{heat}\approx12\,\unit{\kelvin}$ to identify the heat transport regime in channels. The magnitude of $T_\mathrm{heat}(I_\mathrm{heat})$ varies, depending on the electron diffusion regime~\cite{PhysRevB.49.14066,PhysRevB.52.4740,PhysRevB.59.2871,PhysRevB.59.13054}. Without allowing for phonon emission by hot electrons, one expects $T_\mathrm{heat}=T_0/2\left[1+(\nu+\nu^{-1})\mathrm{arctan}(\nu)\right]$~\cite{PhysRevB.59.2871} with $\nu=\sqrt{3}eV_\mathrm{bias}/(2{\pi}k_\mathrm{B}T_0)$, where $V_\mathrm{bias}=I_\mathrm{heat}R_\mathrm{ch}$ is the voltage bias on the heater channel. This formula describes a transition from quadratic heating with $T_\mathrm{heat} \approx T_0(1+\nu^2/2)$ at $eV_\mathrm{bias}{\ll}k_\mathrm{B}T_0$ to a linear temperature increase with $T\approx\frac{\sqrt{3}}{4}\frac{eV_\mathrm{bias}}{2k_\mathrm{B}}$ when $eV_\mathrm{bias}{\gg}k_\mathrm{B}T_0$. In the range up to $T_\mathrm{heat}\sim3\,\unit{\kelvin}$ we rather observe a slower growth of $T_\mathrm{heat}$ (Fig.~\ref{fig:kappa}a). For higher bias, we observe a characteristic bending (Supplemental materials), indicating a phonon contribution to the heat relaxation.

We identify the phonon relaxation law by matching the experiment with a finite element simulation based on the heat transport equation (Supplemental material). We find the electron-phonon relaxation per area in a form $q_\mathrm{ph}=\Sigma_\mathrm{ph}(T^3 - T_0^3)$ with coefficient $\Sigma_\mathrm{ph}\approx0.19\,\unit{\watt\meter^{-2}\kelvin^{-3}}$. The power 3 corresponds to the relaxation via emission of polar acoustic phonons at low temperature~\cite{Ridley_1991}. A relaxation rate with same law has been observed previously for narrow HgTe quantum wells~\cite{PhysRevB.96.245417}.

If electronic diffusion is the sole mechanism of heat transport, one expects:
\begin{equation}
\begin{aligned}
T_\mathrm{heat} &= \frac{1}{2}\left[ T_0 + \frac{T_\mathrm{hot}^2}{\sqrt{T_\mathrm{hot}^2 - T_0^2}}\arccos\frac{T_0}{T_\mathrm{hot}}\right], \\
T_\mathrm{det} &= \frac{2}{3}\left[ T_0 + T_\mathrm{cold} - \frac{T_\mathrm{cold}T_0}{T_\mathrm{cold}+T_0}\right].
\end{aligned}
\end{equation}
The difference between these two expressions comes from the profile for heat injection from a central contact versus uniform Joule heating.

The inclusion of electron-phonon relaxation in the system affects the coefficients in the above equations. Hence, careful consideration of all forms of heat relaxation is essential when deriving these coefficients. Conducting a thorough numerical modeling that incorporates possible electron-phonon relaxation allows for an accurate estimate of the coefficients, deviating slightly from those in a purely diffusive scenario. The relation between $T_\mathrm{heat}$ and $T_\mathrm{hot}$, and between $T_\mathrm{det}$ and $T_\mathrm{cold}$ are also found to be modified and are taken into consideration in the calculation of the thermal conductance (Supplemental material).

When heat transport is exclusively mediated by electron diffusion, the average temperatures $T_\mathrm{heat}$ and $T_\mathrm{det}$ for a rectangular strip can be expressed as follows (in the limit where both $T_\mathrm{heat}-T_0$ and $T_\mathrm{det}-T_0$ are much smaller than $T_0$~\cite{PhysRevB.104.L161403}):
\begin{equation}
		\begin{aligned}
			T_\mathrm{heat} &= (P_\mathrm{J}R_\mathrm{ch})/(12L_0 T_0), \\
			T_\mathrm{det} &= (QR_\mathrm{ch})/(8L_0 T_0)
		\end{aligned}
		\label{eq:heater detector coefficients}
\end{equation}
where $P_\mathrm{J}=I_\mathrm{heat}^2R_\mathrm{ch}$, the dissipated heating power. By assuming $T_\mathrm{heat}=T_\mathrm{det}$ we get the ratio $Q/P_\mathrm{J}=8/12=2/3$. Since the detector temperature only maximally increases by $0.1\,\unit{\kelvin}$ on top of $T_0=1.33\,\unit{\kelvin}$, this is the correct limit to analyze the heat relaxation in the detector channel. We do still use the numerical simulation to refine the value of the coefficients for our device geometry.

\subsection*{Thermovoltage measurement}

For a thermovoltage measurement we pass a low-frequency alternating current through one of the channels, thereby inducing a time-dependent electron temperature difference across the island. The resulting thermal voltage develops at twice the excitation frequency, and is measured using a standard lock-in technique at the second harmonic frequency. This thermal voltage is subsequently converted to the Seebeck coefficient $S$, utilizing the electron temperature difference between the heater and detector channels, which is determined via Johnson noise thermometry in the regime where the temperature has quadratic dependence on the heating current (close to $T_0$).

\subsection*{Acknowledgements}
We thank Mohamed Abdelghany, Yuan Yan and Mitali Banerjee for their help in developing the Johnson-Nyquist thermometry technique, Piotr Sur\'owka, Roderich Moessner and Francisco Pe\~{n}a-Benitez for useful discussions on electron heat transport in Weyl semimetals, and Andrew Mackenzie and Claudia Felser for hosting our Max-Planck Fellowship. This work was supported by the DFG through project SFB 1170 (Project ID 258499086) and the Würzburg-Dresden Cluster of Excellence on Complexity and Topology in Quantum Matter (EXC 2147, Project ID 390858490); by the Free State of Bavaria through the Institute for Topological Insulators; and through a Max Planck fellowship at the Max Planck Institute for Chemical Physics of Solids, Dresden.

\subsection*{Author contributions}
A.A.A., S.U.P., H.B., and L.W.M. planned and designed the experiment. D.C. grew the material, and F.S. fabricated the HgTe device under supervisions of J.K. A.A.A. performed the experiments and carried out the numerical simulations assisted by S.U.P and W.B. Y.J.H contributed in developing the measurement setup for noise thermometry. All authors participated in the analysis led by A.A.A. and S.U.P. All authors participated in writing of the manuscript.


\newpage
\clearpage
\widetext

\setcounter{equation}{0}
\setcounter{figure}{0}
\setcounter{table}{0}
\setcounter{page}{1}
\setcounter{section}{0}

\renewcommand{\refname}{Supplementary References}

\renewcommand{\thefigure}{S\arabic{figure}}
\renewcommand{\theequation}{S\arabic{equation}}
\renewcommand{\thesection}{NOTE \arabic{section}}
\renewcommand{\bibnumfmt}[1]{[S#1]}
\renewcommand{\citenumfont}[1]{S#1}
\renewcommand{\theHfigure}{S\arabic{figure}}

\begin{center}
	\textbf{\large Supplementary Information: Wiedemann-Franz behavior at the Weyl points in compressively strained HgTe}
\end{center}

\section{Johnson Nyquist Noise: Measurement Scheme}

\begin{figure*}[hbt!]
	\centering
	\includegraphics[scale=0.6]{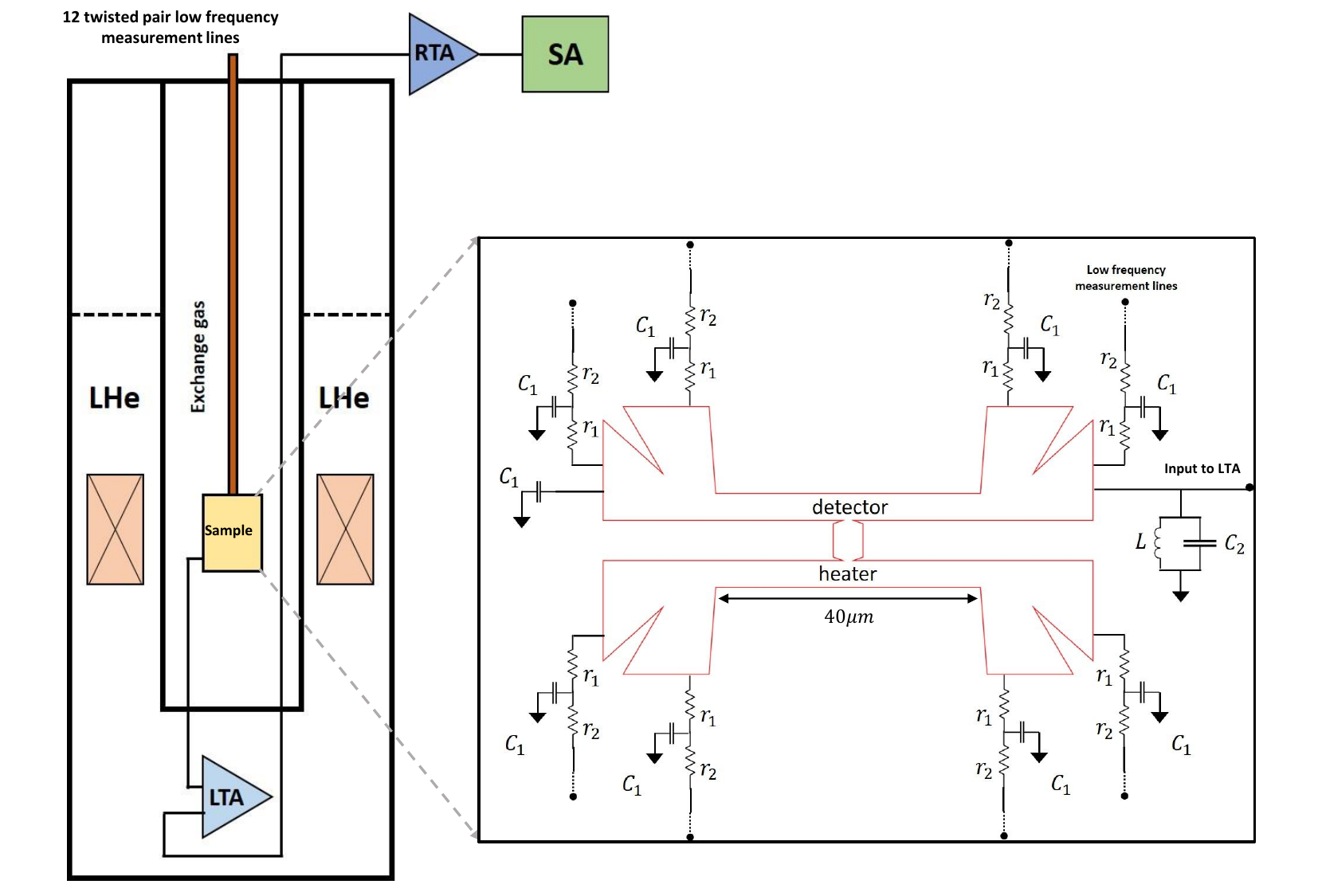}
	\caption{
		Johnson noise measurement setup . 
	}
	\label{fig:Noise setup}
\end{figure*}

Fig.~\ref{fig:Noise setup} shows a schematic diagram of the measurement setup for Johnson-Nyquist noise. A custom-built low-noise, low-temperature amplifier (LTA) first amplifies noise signal from the sample and which is then passed to another amplification stage at room-temperature. A spectrum analyzer (SA) processes the amplified signal at the resonant frequency of the LC tank circuit connected between the sample and LTA. The sample is enclosed in a chamber with a small amount of helium exchange gas. The sample chamber, along with the LTA placed on the outside the chamber, is immersed in the helium bath. The helium bath temperature is maintained at 1.33K through evaporative cooling. The sample is cooled to bath temperature by the exchange gas within the chamber. With this configuration, the sample temperature can be changed by a local heater, while keeping the LTA temperature at constant bath temperature.

The low-frequency lines, used for AC lock-in measurements and application of gate voltages, are fitted with RC low-pass filters that block the outside RF signals from reaching into the device. The resistance in these filters is chosen such that the RF signals from the device are effectively choked and the lines have minimal influence on the high frequency noise measurements. The capacitor \(C_1\) (10 nF) connected to one terminal of the channel acts as a high-frequency ground, which is critical for stabilizing the circuit at higher frequencies. The other terminal is connected to the input of the LTA via an LC tank. The resistor \(r_2\) (1 k\(\Omega\)) and capacitor \(C_1\) form a low-pass RC filter with a cutoff frequency of 16 kHz. These filters attenuate unwanted high frequency noise and enhance the signal-to noise ratio. The resistor \(r_1\) controls the noise distribution along the sample and isolates the detector side from other components of the H-bar structure. The value of \(r_1\) is optimized to prevent excessively high resistance, which could limit the frequency range of low-frequency AC measurements. Based on our calculations, \(r_1\) was set to 50 k\(\Omega\), ensuring that only 0.26\% of thermal noise from other parts of the H-bar structure reaches the detector. Moreover, a home-built air-core superconducting inductor with an inductance of L=$ 10\,\unit{\micro H}$ is employed to form an LC tank circuit for impedance matching. The system’s parasitic capacitance, including the coaxial cable capacitance (approximately C\textsubscript{2}=$78\,\unit{\pico F}$), results in a resonant frequency of f\textsubscript{res}=$5.7\,\unit{\MHz}$.

\section{Temperature Profile across Heater and Detector Channel}

The zero \(B\)-field temperature profiles in the heater and detector channels were simulated using finite element analysis, accurately adapting the device geometry shown in the inset of Fig.~\ref{fig:temperature profile}.d. Figures Fig.~\ref{fig:temperature profile}.a and Fig.~\ref{fig:temperature profile}.b depict the temperature profiles along the center of the heater (black) and detector (red) channels under different heating excitations of $200\,\unit{\micro\volt}$ and 2 mV, respectively. The heater exhibits a parabolic temperature profile due to uniform Joule heating, while the detector shows a linear profile, because the heat is injected in the middle of the detector channel\cite{S_PhysRevB.104.L161403}. Notably, large contact areas at both ends of the channels significantly alter the temperature profile compared to the regions within the channels. Additionally, the temperature profiles in both the heater and detector channel are affected by the level of heating. Fig.~\ref{fig:temperature profile}.c illustrates the correlation between the average temperature in the heater (\(T_{\text{heat}}\)) and detector (\(T_{\text{det}}\)) channels and the local temperatures inside the channels close to the island entrance (\(T_{\text{hot}}\) for the heater and \(T_{\text{cold}}\) for the detector). Excitation-dependent conversion factors derived from this analysis are used to translate the experimentally measured \(T_{\text{heat}}\) and \(T_{\text{det}}\) into \(T_{\text{hot}}\) and \(T_{\text{cold}}\). The slight modifications in the temperature profiles of the heater and detector are also reflected in the small changes in the ratio of heat flow across the island to the dissipated heating power in the channel, \(Q/P_J\). This ratio is also dependent on the heating excitation and was found to deviate slightly from the theoretically predicted value of 2/3 (Fig.~\ref{fig:temperature profile}.d). A similar analysis was conducted for all other \(B\)-field values as well. 

\section{Heat Relaxation to Phonons in Different Transport Regime}

By varying the top gate voltage of the heater channel, \(T_{\text{heat}}\) is measured for different Fermi level positions at \(B = 0 \, \text{T}\): the electron regime, the Weyl point, and the hole regime, as shown in the scatter plots in Fig.~\ref{fig:phonon relaxation}. A characteristic bending, indicated by a star symbol, is observed in the higher heating range, which suggests the onset of a phonon contribution to heat relaxation. The \(T_{\text{heat}}\) values for these scenarios were simulated (solid line plots in Fig.~\ref{fig:phonon relaxation}) using finite element analysis based on the heat transport equation, incorporating charge carrier-phonon relaxation in the form \(q_{\text{ph}} = \Sigma_{\text{ph}} (T^3 - T_0^3)\). The value of \(\Sigma_{\text{ph}}\) in the electron regime and at the Weyl point is nearly identical (0.19 Wm\(^{-2}\)K\(^{-3}\)), while in the hole regime, \(\Sigma_{\text{ph}}\) is approximately double (0.36 Wm\(^{-2}\)K\(^{-3}\)), indicating enhanced hole-phonon scattering , likely due to the larger effective mass of the holes.

\section{Heat Relaxation Equivalence in Heater and Detector Channels}

The LTA is always connected to the same channel. When \(I_{\text{heat}}\) is applied to the channel with the LTA, the increase in electron temperature, \(T_{\text{heat}}\), is measured. When \(I_{\text{heat}}\) is applied to the other channel, the LTA measures \(T_{\text{det}}\). To ensure identical heat relaxation mechanisms in both the heater and detector channel, additional measurements are taken after reconnecting the LTA to the other channel. Despite requiring a thermal cycle for connection, no significant changes in gate-dependent resistance of the channels were observed. Fig.~\ref{fig:heat relaxation equivalance} shows the electron temperature response to \(I_{\text{heat}}\) for both channels, with the curves matching well, confirming the validity of the assumption that both channels exhibit the same heat relaxation behavior.

\section{Error Analysis}

Both the X and Y axes in Fig. 3b contribute to the uncertainty in the thermal conductance, \(\kappa\), where \(X = (T_{\text{hot}}^2 - T_{\text{cold}}^2)\) and \(Y=Q \approx (2/3) P_J(T_{heat}=T_{det})\). The uncertainties in \(X\) and \(Y\), denoted \(\Delta X\) and \(\Delta Y\), are calculated using standard error propagation for functions of multiple variables, as follows:

$	\Delta X = \sqrt{ \left( \frac{\partial X}{\partial T_{\text{hot}}} \right)^2 (\Delta T_{\text{hot}})^2 + \left( \frac{\partial X}{\partial T_{\text{cold}}} \right)^2 (\Delta T_{\text{cold}})^2 }$

$	\Delta X = 2 \sqrt{T_{\text{hot}}^2 (\Delta T_{\text{hot}})^2 + T_{\text{cold}}^2 (\Delta T_{\text{cold}})^2 }$

$\Delta Y = \sqrt{ \left( \frac{\partial Y}{\partial T_{\text{heat}}} \right)^2 (\Delta T_{\text{heat}})^2 + \left( \frac{\partial Y}{\partial T_{\text{det}}} \right)^2 (\Delta T_{\text{det}})^2 }$

The uncertainty in the slope of the \(X\)-\(Y\) plot, corresponding to \(\kappa\), is then determined by performing a weighted linear regression, with the relevant expressions provided below:

$\Delta \kappa = \sqrt{\frac{1}{S_{xx}}}$ where

$S_{xx} = \sum_{i=1}^n w_i X_i^2$ and the effective weight, $\quad w_i = \frac{1}{\Delta Y_i^2 + \kappa^2 \Delta X_i^2}$

The final error bars in the main graph represent a 95\% confidence interval, ensuring robust statistical significance.

\begin{figure*}[hbt!]
	\centering
	\includegraphics[scale=1.4]{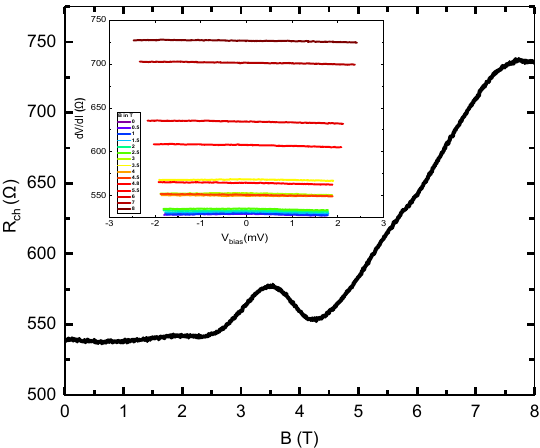}
	\caption{In-plane magneto-resistance of the heater channel when the channel fermi level tuned to the n-conducting regime (n $\approx$ 3$\times$10\textsuperscript{11}cm\textsuperscript{-2}). The inset shows the differential resistance of the heater for different in-plane B fields, ranging from 0 to 8T}
	\label{fig:channel magnetoresistance}
\end{figure*}

\begin{figure*}[hbt!]
	\centering
	\includegraphics[scale=0.95]{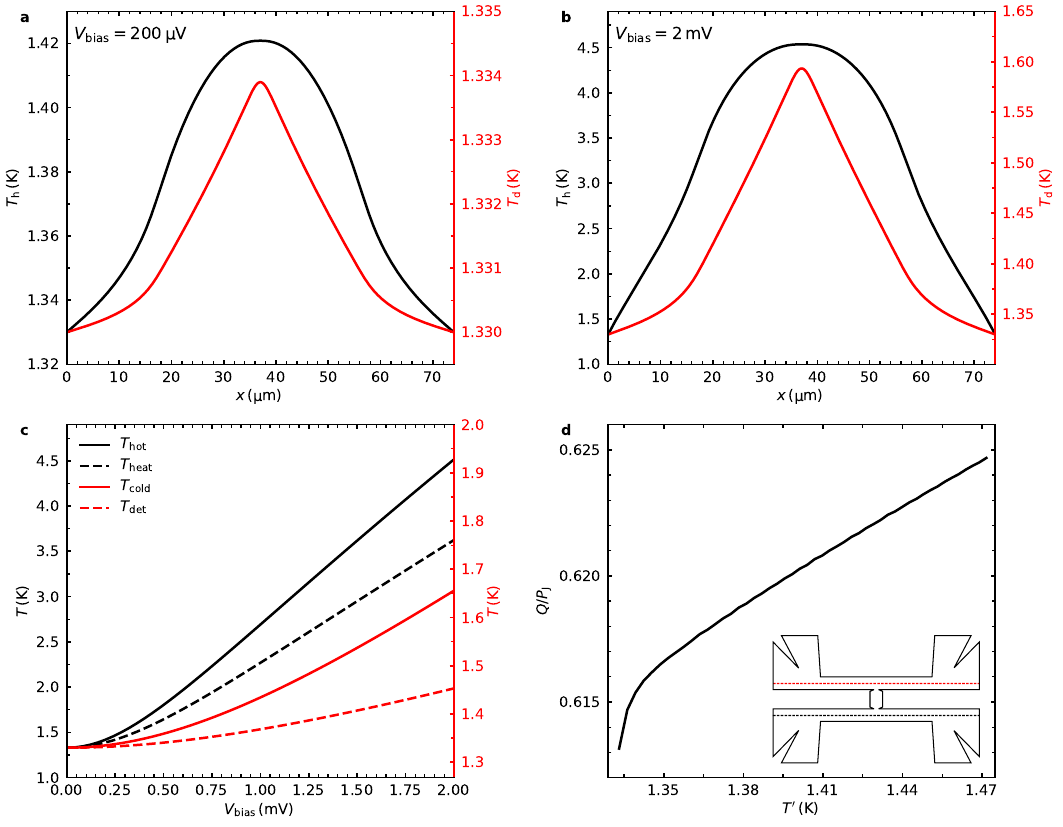}
	\caption{Temperature profile across the centre of the heater (a) and detector (b) channels. (c) Average temperature over the channels (\(T_{heat}\) and \(T_{det}\))and local temperature (\(T_{hot}\) and \(T_{cold}\)) close to the island entrance for different heating excitations. (d) Temperature dependent ratio of heat flow across the island to the dissipated heating power in the channel, \(Q/P_J\).The inset illustrates the layout of the sample design used for the simulation. The black dashed line corresponds to the line scan along the heater, while the red dashed line indicates the scan along the detector.}
	\label{fig:temperature profile}
\end{figure*}

\begin{figure*}[hbt!]
	\centering
	\includegraphics[scale=1.4]{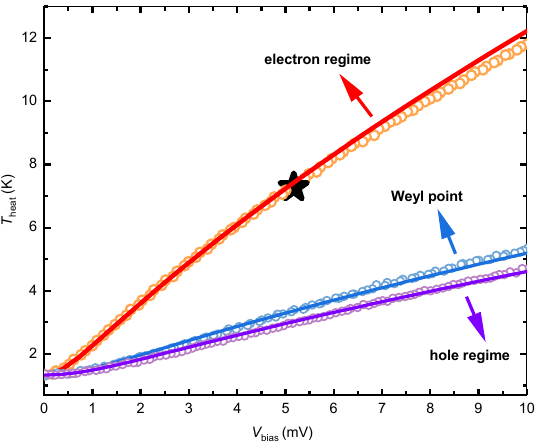}
	\caption{Excitation dependent  electron temperature in the heater channel, \(T_{heat}\) for different Fermi level positions: electron regime (red), Weyl point (blue) and hole regime (violet). Solid lines corresponds to the simulations while scattered plots represents the experimentally measured data.}
	\label{fig:phonon relaxation}
\end{figure*}

\begin{figure*}[hbt!]
	\centering
	\includegraphics[scale=1.4]{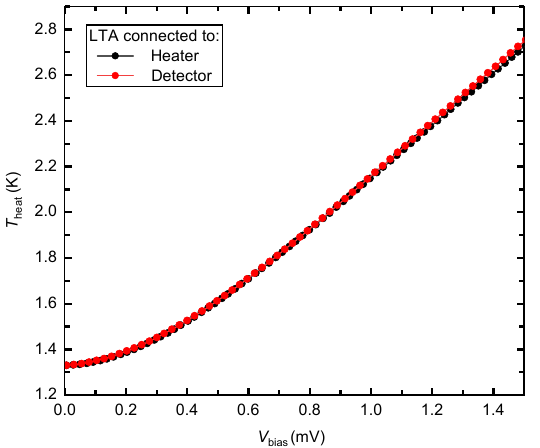}
	\caption{Electron temperature response to \(I_{\text{heat}}\) in both the heater and detector channels, measured using the LTA. The close match between the curves confirms the assumption of identical heat relaxation mechanisms in both channels, even after the LTA reconnection.}
	\label{fig:heat relaxation equivalance}
\end{figure*}

\end{document}